\documentclass[twoside]{photon2007}
\usepackage[latin1]{inputenc}
\usepackage[dvips]{graphicx,epsfig,color}
\usepackage{wrapfig,rotating}
\usepackage{amssymb,amsmath,array}

\newcommand{\reff}[1]{(\ref{#1})}
\begin{document}
\title{
The alternative approach to QCD analysis of the structure function
$F_2^{\gamma}$ } 
\author{Jiri Hejbal
\vspace{.3cm}\\
Institute of Physics of the Academy of Sciences, Czech Republic \\
\vspace{.1cm}\\
}

\maketitle

\begin{abstract}
The alternative approach to QCD analysis of the structure function
$F_2^{\gamma}$ is presented. It differs from the conventional one by the
presence of the terms, which in conventional approach appear in higher
orders. The numerical results show that this difference is nonnegligible
and may play an important role in the analysis on photon data of the
future experiments. 
\end{abstract}

\section{Parton distribution functions of the photon and their evolutions}
Distribution functions $q^{\gamma}(x,Q^2)$ of the photon satisfy the inhomogeneous evolution equations
\begin{eqnarray}
\frac{\textup{d} \Sigma^{\gamma}(x,Q^2)}{\textup{d}\ln Q^2} &\!\!\!\!\!\!\!=&\!\!\!\!\!\!  \delta_{\Sigma}k_q+ P_{qq}\otimes \Sigma^{\gamma}+P_{qG}\otimes G^{\gamma}\nonumber\\
\frac{\textup{d} G^{\gamma}(x,Q^2)}{\textup{d}\ln Q^2} &\!\!\!\!\!=&\!\!\!\!\!  k_G+ P_{Gq}\otimes \Sigma^{\gamma}+P_{GG}\otimes G^{\gamma}\nonumber\\
\frac{\textup{d} q^{\gamma}_{NS}(x,Q^2)}{\textup{d}\ln Q^2} &\!\!\!\!\!=& \!\!\!\!\! \delta_{NS}k_q+ P_{qq}\otimes q^{\gamma}_{NS}
\label{PDF}
\end{eqnarray}
where the $\otimes$ product denotes the conventional convolution in x-space.

Here
\begin{eqnarray}
&&\Sigma^\gamma(x,Q^2)=\sum_f[q^\gamma_f(x,Q^2)+\bar{q}^\gamma_f(x,Q^2)],\nonumber\\
\lefteqn{q^\gamma_{NS}(x,Q^2)=}\nonumber\\
&&\sum_f(e_i^2-\langle e^2 \rangle^2 )(q^\gamma_f(x,Q^2)+\bar{q}^\gamma_f(x,Q^2)),\nonumber
\end{eqnarray}
where \emph{f} runs over all relevant quark flavors.

To order $\alpha$ the photon-parton splitting functions $k(x,Q^2)$ and the parton-parton splitting functions $P(x,Q^2)$ in Eq. $\reff{PDF}$ are given as power expansions in $\alpha_S(Q^2)$.

\section{Solving evolution equations}

The evolution equations $\reff{PDF}$ in Mellin moments
\begin{equation}
 q_i^\gamma(n,Q^2) \equiv \int_0^1 \textup{d}xx^{n-1}q_i^\gamma(x,Q^2),
\end{equation}
read
\begin{eqnarray}
\frac{\textup{d} \Sigma^{\gamma}(n,Q^2)}{\textup{d}\ln Q^2} &\!\!\!\!\!=&\!\!\!\!\!  \delta_{\Sigma}k_q+ P_{qq} \Sigma^{\gamma}+P_{qG}G^{\gamma}\nonumber\\
\frac{\textup{d} G^{\gamma}(n,Q^2)}{\textup{d}\ln Q^2} &\!\!\!\!\!=&\!\!\!\!\!  k_G+ P_{Gq}\otimes \Sigma^{\gamma}+P_{GG} G^{\gamma}\nonumber\\
\frac{\textup{d} q^{\gamma}_{NS}(n,Q^2)}{\textup{d}\ln Q^2} &\!\!\!\!\!=&\!\!\!\!\! \delta_{NS}k_q+ P_{qq} q^{\gamma}_{NS}\nonumber
\label{PDFn}
\end{eqnarray}
The solution of \reff{PDF} can be separated into the so-called pointlike part, related to a special solution of the full inhomogeneous equation and hadron-like part, arising as a general solution of the homogeneous equation. In Mellin moments it reads:
\begin{equation}
q^\gamma(n,Q^2)=q^\gamma_{PL}(n,Q^2)+q^\gamma_{had}(n,Q^2).
\label{PL&HAD}
\end{equation}
 It is straightforward to find such a solution in Mellin moments. The LO pointlike solution is given by \cite{GRV1}
\begin{eqnarray}
q^{\gamma}_{PL} = \frac{4 \pi} {\alpha_S(Q^2)} \left[1-\left(\frac{\alpha_S(Q^2)}{\alpha_S(Q^2_0)}\right)^{1-(\frac{2}{\beta_0})P^{(0)}}\right] \nonumber\\
\frac{1}{1-\frac{2}{\beta_0}P^{(0)}(n)}\frac{\alpha}{2\pi\beta_0}k^{(0)}(n) \nonumber
\label{PL}
\end{eqnarray}
and the LO hadronic solution by
\begin{eqnarray}
q_{had}^{\gamma}(Q) &=&  \left(\frac{\alpha_S(Q^2)}{\alpha_S(Q^2_0)}\right)^{-2P^{(0)}/\beta_0} q_{had}^{\gamma}(Q_0^2) \nonumber
\label{HAD}
\end{eqnarray}

With the explicit solutions for the photonic parton distributions $q_{NS}^{\gamma}$, $\Sigma^{\gamma}$ and $G^{\gamma}$ at hand it is now straightforward to obtain the photon structure function $F_2^{\gamma}(n,Q^2)$ given as
\begin{eqnarray}
 \lefteqn{F_2^{\gamma}(n,Q^2)=}\nonumber\\
&\!\!\!\!\!\!\!\!\!\!\!\!&\!\!\!\!\!\!\!\!\!\! q^{\gamma}_{NS}(n,Q^2)C_q(n)+\frac{\alpha}{2\pi}\delta_{NS}C_{\gamma}(n)+ \nonumber\\
&\!\!\!\!\!\!\!\!\!\!\!\!&\!\!\!\!\!\!\!\!\!\! <\!\!e^2\!\!>\left(\Sigma(n,Q^2)C_q(n)+G(n,Q^2)C_G(n)\right)+\nonumber\\
&\!\!\!\!\!\!\!\!\!\!\!\!&\!\!\!\!\!\!\!\!\!\! \frac{\alpha}{2\pi}<e^2>\delta_{\Sigma}C_{\gamma}(n)+\nonumber
\label{F2gamma}
\end{eqnarray}
where the coefficient functions $C_q(x)$, $C_G(x)$ and $C_{\gamma}(x)$ enter the $F_2^{\gamma}$ in convolution with the photon distributions.

In the following chapter I will present a basic features of the alternative approach to the analysis of the structure function $F_2^{\gamma}$.

\section{Alternative approach to QCD analysis of $F_2^{\gamma}$}

First I will recall the conventional formulation of QCD analysis of $F_2^{\gamma}$ in LO. This approach is based on the assumption that

\begin{eqnarray} \frac{1}{x}F_{2}^{\gamma,LO}(x,Q^2)=q_{NS}^{\gamma}(x,Q^2)+ \nonumber\\
<e^2> \Sigma^{\gamma}(x,Q^2) \label{F2conv}\nonumber
\end{eqnarray}
with the pointlike parts of non-singlet and singlet distributions satisfing the evolution equation with r.h.s including splitting functions $k^{(0)}$ and $P^{(0)}$ only.

The alternative approach \cite{Chyla} is based on different definition of the terms "leading" and the "next-to-leading" approximation to $F_2^{\gamma}$. It is argued \cite{Chyla} that complete LO analysis of the $F_2^{\gamma}$ requires the inclusion of four known, but in the conventional LO analysis unused quantities. Note, that the modification concerns only point-like part of $F_2^{\gamma}$. For more details, please see \cite{Chyla}.

The LO formula for $F_2^{\gamma,PL}$ in this approach reads
\begin{eqnarray}
\lefteqn{ \frac{1}{x}F_{2,alter.}^{\gamma,LO,PL}(x,Q^2)=}\nonumber \\
&&q_{NS}^{\gamma,PL}(x,Q^2)+\left\langle e^2 \right\rangle  \Sigma_{PL}^{\gamma}(x,Q^2)+\nonumber \\
&&\frac{\alpha_s}{2\pi}C_{q}^{(1)}\otimes q^{\gamma}_{NS}+\left\langle e^2\right\rangle\frac{\alpha_s}{2\pi}C_{q}^{(1)}\otimes q^{\gamma}_{\Sigma}+ \nonumber \\
&&\delta_{NS}\frac{\alpha}{2\pi}C_{\gamma}^{(0)}+\left\langle e^2\right\rangle\delta_{\Sigma}\frac{\alpha}{2\pi}C_{\gamma}^{(0)}+\nonumber \\
&&\delta_{NS}\frac{\alpha}{2\pi}\frac{\alpha_s}{2\pi}C_{\gamma}^{(1)}+\left\langle e^2\right\rangle\delta_{\Sigma}\frac{\alpha}{2\pi}\frac{\alpha_s}{2\pi}C_{\gamma}^{(1)}. \nonumber
\end{eqnarray}

This expression differs from $\reff{F2conv}$
\begin{itemize}
\item 
by appearance of the contributions of the photonic coefficient functions $C_{\gamma}^{(0)}$ and $C_{\gamma}^{(1)}$,
\item
by appearance of the convolution of quark coefficient function $C_q^{(1)}$ with $q_{NS}$ and $q_{\Sigma}$,
\item
by the fact that $k_{NS}^{(1)}$ resp. $k_q^{(1)}$ and $k_g^{(1)}$ are included in the evolution equations for $q_{NS}^{\gamma}$ resp. $\Sigma^{\gamma}$.
\end{itemize}

\begin{figure}[ht]
\centerline{\includegraphics[width=1.0\columnwidth]{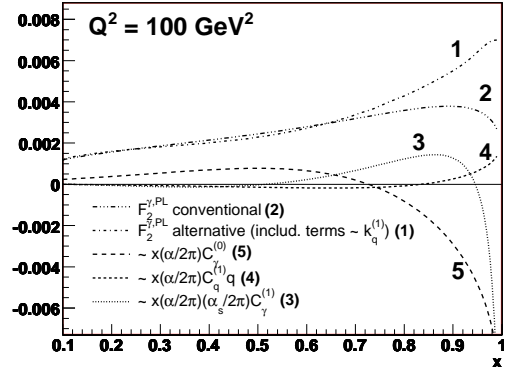}}
\caption{Contributions to LO QCD expression for the pointlike part of $F_{2}^{\gamma}$ included in alternative approach for $Q^2=100\  \textup{GeV}^2$}\label{Fig:PH1}
\end{figure}

The numerical representation of all contributions are shown in Fig.\reff{Fig:PH1}. Note, that all contributions evolve in $Q^2$ (except the term proportional to $C_{\gamma}^{(0)}$, which doesn't depend on $Q^2$), but the shapes of these contributions stay unchanged (as shown on animated graphs in \cite{url}). The common contribution of terms proportional to the coefficient functions $C_{\gamma}^{(0)}$ and $C_{\gamma}^{(1)}$ give the positive contribution in the region $x\lesssim 0.85$. In region $x\gtrsim0.85$ dominates the negative contribution $\propto C_{\gamma}^{(0)}$, which is strengthened by negative contribution $\propto C_{\gamma}^{(1)}$ close to $x= 1$.

\begin{figure}[ht]
\centerline{\includegraphics[width=1.0\columnwidth]{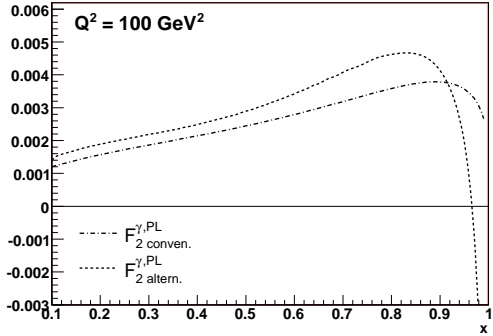}}
\caption{Comparison of the pointlike part of $F_2^{\gamma}$ in conventional and alternative approach for $Q^2=100\  \textup{GeV}^2$}\label{Fig:PH2}
\end{figure}

The contribution $\propto C_{q}^{(1)}$, entering throught the convolution with distribution function $q_{NS}$ and $q_{\Sigma}$ has quantatively the same shape as $k_{q}^{(1)}$ and make the positive corection close to $x= 1$.

Finally, the effect of taking the $k_q$ into the evolution equations for $q$ is evident from comparison of line 1 and 2 in Fig. \reff{Fig:PH1}. 

Putting contributions together we can compare $F_{2,PL}^{\gamma}$ in both approaches (Fig. \reff{Fig:PH2}). Numerically, in the alternative approach, $F_{2,PL}^{\gamma}$ lie higher in region up to $x\doteq 0.9$, in region $x\doteq 0.8$ even notably higher. In contrary, in the region close to $x= 1$, the values of $F_{2,PL}^{\gamma}$ in alternative approach decrease much faster then values of $F_{2,PL}^{\gamma}$ in convetional one. This is due to the inclusion of $C^{(0)}_{\gamma}$.

\begin{figure}[ht]
\centerline{\includegraphics[width=1.0\columnwidth]{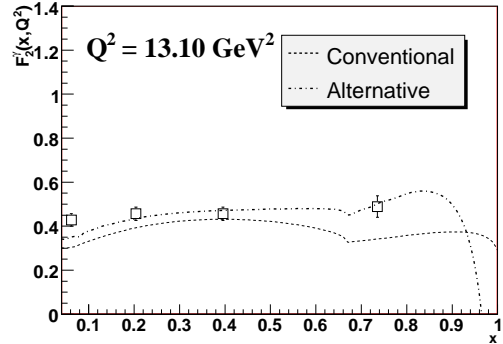}}
\caption{Example of comparison of the numerical difference of two approaches of counting $F_2^{\gamma}$ in LO with the experimental errors for $Q^2=13.1\ \textup{GeV}^2$}\label{Fig:PH3}
\end{figure}

The significance of numerical difference between both approaches becomes more clear, if we compare these differences with errors of data on $F_{2}^{\gamma}$. If we aim to the data where their errors are resonably small, we can conclude, that this difference is numerically important. An ilustration is presented in Fig. \reff{Fig:PH3}. Is evident that the numerical difference between both approaches is bigger then the accuracy of measurement of data on $F_{2}^{\gamma}$. Another exaple could be found in \cite{url}.

\section{QCD analysis of the structure function $F_2^{\gamma}$ in LO}

In this chapter I will briefly present the results of the QCD analysis of the structure function $F_2^{\gamma}$ in alternative approach. I adopted the Fixed-Flavour-Number Scheme model ($FFNS_{CJKL}$) presented in \cite{kraw} as a toy model, where the implementation of the alternative approach will be easy to realize and I compared my results with those of autors of \cite{kraw}.

In $FFNS_{CJKL}$ we deal only with 3 lights quarks (u,d,s), contributions of the massive $c$ and $b$ quarks are assumed to be given by the Bethe-Heitler formula. The input scale is chosen to be small, $Q^2_0=0.25\ \textup{GeV}^2$ and the input hadronic distribution generated via Vector Meson Dominance (VDM) model. For more details see \cite{kraw}.

\subsection{Global fits and results}


I have performed fit to 182 data points. It is based on the least-squares principle (minimum of $\chi^2$) performed in MINUIT. Systematic and statistical errors on data points were added in quadrature.

For $n_f=3$ I took the QCD scale $\Lambda^{(3)}=314$ MeV and the masses of the heavy quarks are taken to be $m_b=1.3$ GeV and $m_c=4.3$ GeV 

\begin{figure}[ht]
\centerline{\includegraphics[width=1.0\columnwidth]{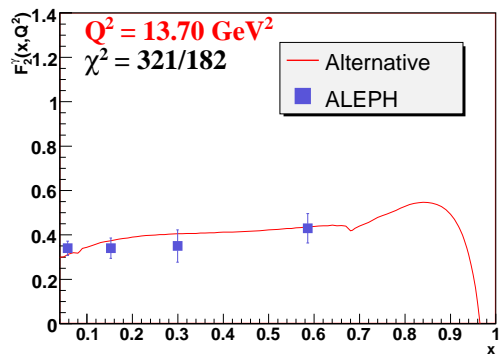}}
\caption{Example of fit performed for the structure function $F_2^{\gamma}$ in alternative approach for $Q^2=13.7\ \textup{GeV}^2$. For 182 data points I reached $\chi^2=321$}\label{Fig:PH4}
\end{figure}

Three parameters $\alpha, \beta$ and $\kappa$ describing a boundary condition and value of $\chi^{2}$ in both approaches are presented in Tab.\reff{srr1}. In alternative approach I have reached a slightly smaller value of $\chi^2$ then in conventional one presented in \cite{kraw}. Graphical ilustration of the result of my fit is in Fig. \reff{Fig:PH4}. For full set of figures, please see \cite{url}.
\begin{table}[ht]
\begin{center}
\begin{tabular}{  | c | c | c | c | c | }
\hline
      \bf Approach & \bf $\chi^2$ &\bf $\kappa$ &\bf $\alpha$ &\bf $\beta$  \\
\hline
   Coventional  & 357  & 1.726 & 0.465 & 0.127 \\
\hline
   Alternative  & 321 & 0.899 & 1.236 & 3.103 \\
\hline
\end{tabular}
\end{center}
\caption{The $\chi^2$ for 182 points and parameters of the fit in conventional and alternative approach in the $FFNS_{CJKL}$ model.}
\label{srr1}
\end{table}

At present, there is no obstacle to perform NLO analysis of $F_2^{\gamma}$ in alternative approach. Such an analysis requires the inclusion of quantities $k^{(2)},\ C_{\gamma}^{(1)}$ and $C_{\gamma}^{(2)}$ in addition to the conventional one. All these quantities have already been calculated or can be derived from those of \cite{VAM}. The work on this is currently in progress.

\begin{footnotesize}



%

\end{footnotesize}


\end{document}